# Experimental Powder X-ray Diffraction Crystal Structure Determination with RealPXRD-Solver

*Qi Li[1,3], Mingyu Guo[1,8], Rui Jiao[9], Jing Gao[1,2,4], Fanjie Xu[1,12], Haonan Xue[1,13], Weixiong Zhang[8], Wenbing Huang[10], Junchi Yan[4], Linfeng Zhang[1,6], Cheng Wang[1,11], Zhuang Yan[1,5,6,7]*, Guolin Ke[1,6]*, Weinan E[6,7], Zhiyong Tang[5]*, Shifeng Jin[3]*, Lin Yao[1,2,4]**

[1]DP Technology; Beijing, 100080, China.

[2]Zhongguancun Academy; Beijing, 100097, China.

[3]The Beijing National Laboratory for Condensed Matter Physics, Institute of Physics, Chinese Academy of Sciences; Beijing, 100190, China.

[4]Shanghai Jiao Tong University; Shanghai, 200240, China.

[5]National Center for Nanoscience and Technology; Beijing, 100190, China.

[6]AI for Science Institute; Beijing, 100084, China.

[7]Center for Machine Learning Research, Peking University; Beijing, 100871, China.

[8]School of Chemistry, Sun Yat-sen University; Guangzhou, 510275, China.

[9]Institute for AIR, Tsinghua University; Beijing, 100084, China.

[10]Gaoling School of Artificial Intelligence, Renmin University of China; Beijing, 100872, China.

[11]College of Chemistry and Chemical Engineering, Xiamen University; Xiamen, 361005, China.

[12]Institute of Artificial Intelligence, Xiamen University; Xiamen, 361005, China.

[13]School of Electronic Engineering and Computer Science, Peking University; Beijing, 100871, China.

*Corresponding authors. Emails: yanz@pku.edu.cn; kegl@dp.tech; zytang@nanoctr.cn; shifengjin@iphy.ac.cn; yaol@dp.tech

## Abstract

Determining crystal structures directly from experimental powder X-ray diffraction (PXRD) data remains a long-standing challenge, as peak overlap, preferred orientation, and impurities obscure the underlying atomic arrangement. Recent artificial intelligence (AI) models can recover structures from simulated diffraction patterns, but their accuracy degrades sharply on real experimental data, revealing a persistent simulation-to-reality gap. Here we present RealPXRD-Solver, a flow-based generative model trained on 6,250,238 theoretical crystal structures with experiment-mimicking data augmentation. A universal XRD encoder extracts invariant d-spacing–intensity (d–I) fingerprints from both simulated and experimental patterns, providing a shared representation across domains and enabling both lattice-conditioned and lattice-free inference. On a 10,000-structure theoretical benchmark, RealPXRD-Solver achieves a 98.3% Top-20 match rate. Crucially, it transfers robustly to experimental data, attaining Top-1/Top-20 accuracies of 77.9%/91.9% on the CNRS dataset and 78.8%/92.9% on the RRUFF benchmark. Even on structurally non-overlapping subsets that exclude any symmetry-equivalent counterparts from the training corpus, the model maintains Top-20 match rates of 87.0% (CNRS) and 85.7% (RRUFF), indicating genuine generalization beyond memorizing known structures. RealPXRD-Solver reliably handles noisy, textured, and multiphase patterns, and has enabled the automated solution of 39 previously unreported structures from the Powder Diffraction File. This framework effectively bridges theoretical modeling and experimental practice, facilitating rapid, autonomous crystal-structure determination from routine PXRD measurements.

# 1. Introduction

The atomic arrangement within a crystal dictates its physical and chemical properties, making structure determination a cornerstone of materials science[1]. Powder X-ray diffraction (PXRD) is widely used for this purpose because it is fast, accessible, and non-destructive[2]. However, traditional PXRD analysis methods, such as Rietveld refinement[3], depend on a known structural model and struggle to solve unknown crystal structures from first principles. This fundamental limitation arises because three-dimensional reciprocal-space information is irreversibly projected onto one-dimensional diffraction profiles. As materials discovery increasingly shifts toward high-throughput and autonomous experimentation, structure solution methods must therefore become automated, accurate, and robust under realistic experimental conditions.

Recent advances in generative artificial intelligence have opened new possibilities for addressing this long-standing challenge[4-6]. Machine-learning approaches have achieved notable progress in individual PXRD-related tasks, including space-group classification[7-10], lattice-parameter prediction[11, 12], and phase identification[13-16]. More recently, end-to-end generative models—such as PXRDnet[17], Crystalyze[18], DiffractGPT[19], XtalNet[20], and PXRDGen[21]—have demonstrated that complete crystal structures can, in principle, be predicted directly from PXRD patterns with high accuracy, primarily on simulated datasets. Collectively, these works have reframed PXRD-based structure determination as a generative modeling problem, providing influential benchmarks and proof-of-concept demonstrations for the field.

Despite this progress, a critical simulation-to-reality gap continues to limit practical applicability. Experimental PXRD data are intrinsically heterogeneous, containing background contributions from air scattering or sample environments, peak broadening due to finite crystallite size or microstrain, preferred orientation induced by sample preparation, and interference from impurity phases. In addition, diffraction patterns vary substantially across laboratory and synchrotron instruments as a function of wavelength, step size, and instrumental broadening (FWHM), factors that are difficult to exhaustively model in synthetic training data. As a result, current AI models often fail to generalize beyond idealized simulations. Recent autonomous materials discovery platforms that combine robotic synthesis with AI-driven PXRD analysis have exposed these limitations: for example, a high-throughput discovery study reported 43 new materials, yet subsequent analysis revealed significant failure modes in its automated Rietveld refinement pipeline, underscoring the fragility of current approaches when confronted with real experimental data[22].

Beyond experimental noise, this gap also reflects a deeper mismatch between how most AI models formulate the structure-solution problem and how crystallography is performed in practice. In real-world workflows, indexing (determining the unit cell) and structure solution (determining the atomic arrangement) are treated as distinct steps, often carried out by different algorithms and with different confidence levels. Indexing may succeed reliably for many samples, yet fail entirely for others due to peak overlap, preferred orientation, or limited angular range. By contrast, most existing end-to-end AI solvers implicitly assume that both lattice parameters and atomic positions must be inferred jointly from PXRD, even when partial crystallographic information is already available. This conflation not only complicates learning, but also limits flexibility in realistic experimental settings.

To address these challenges, we introduce RealPXRD-Solver, a generative model designed for robust and scalable crystal structure determination directly from experimental PXRD data. RealPXRD-Solver explicitly aligns with practical crystallographic workflows by supporting both a lattice-conditioned setting, where independently determined unit-cell parameters are provided when indexing succeeds, and a lattice-free *ab initio* setting for cases where indexing fails or unit-cell information is unavailable. This unified formulation allows the model to leverage reliable prior information when available, without sacrificing generality. RealPXRD-Solver bridges the simulation-to-reality gap through three key components: (i) a universal XRD encoder that extracts intrinsic, discrete d–I (interplanar spacing–intensity) representations that are invariant to measurement artifacts; (ii) a large-scale training corpus comprising over six million theoretical crystal structures spanning all crystal systems and 89 chemical elements; and (iii) extensive physics-informed data augmentation that emulates realistic experimental perturbations, including peak shifts, broadening, intensity distortion, and noise. Together, these elements enable robust generalization across instruments, sample conditions, and chemical compositions.

RealPXRD-Solver achieves a 98.3% (Top-20) match rate on a large theoretical benchmark. On experimental datasets, it attains 77.9%/91.9% on the CNRS experimental PXRD dataset[23] and 78.8%/92.9% (Top-1/Top-20) accuracy on the RRUFF mineral PXRD dataset[24], both comprising real measured diffraction patterns. The model accurately reconstructs structures even under high noise, preferred orientation, or impurity phases, and can distinguish neighboring elements as well as light atoms. Finally, it demonstrates practical utility by automatically solving 39 previously unreported structures from the Powder Diffraction File (PDF) database, establishing an end-to-end pipeline for automated, high-throughput materials characterization.

## 2. Results and Discussion

### 2.1 Design of RealPXRD-Solver for experimental data

To address the experimental heterogeneity and workflow-related challenges discussed above, we designed RealPXRD-Solver around two key ideas (**Fig. 1**): (i) representing diffraction patterns by an intrinsic, discrete *d–I* (interplanar spacing–intensity) fingerprint, and (ii) processing both simulated and experimental inputs through a unified generative workflow. Unlike models that operate directly on continuous intensity–$2\theta$ profiles, which are highly sensitive to background level, peak broadening, $2\theta$ range, and step size, RealPXRD-Solver first converts raw patterns into *d–I* lists. As illustrated for $LiFePO_4$, a widely used olivine-type Li-ion battery cathode with well-characterized PXRD signatures, in **Fig. 1a** and **Supplementary Fig. 1**, diffraction profiles measured under different conditions collapse onto a nearly identical *d–I* representation, demonstrating that this fingerprint is largely invariant to experimental settings and sample quality.

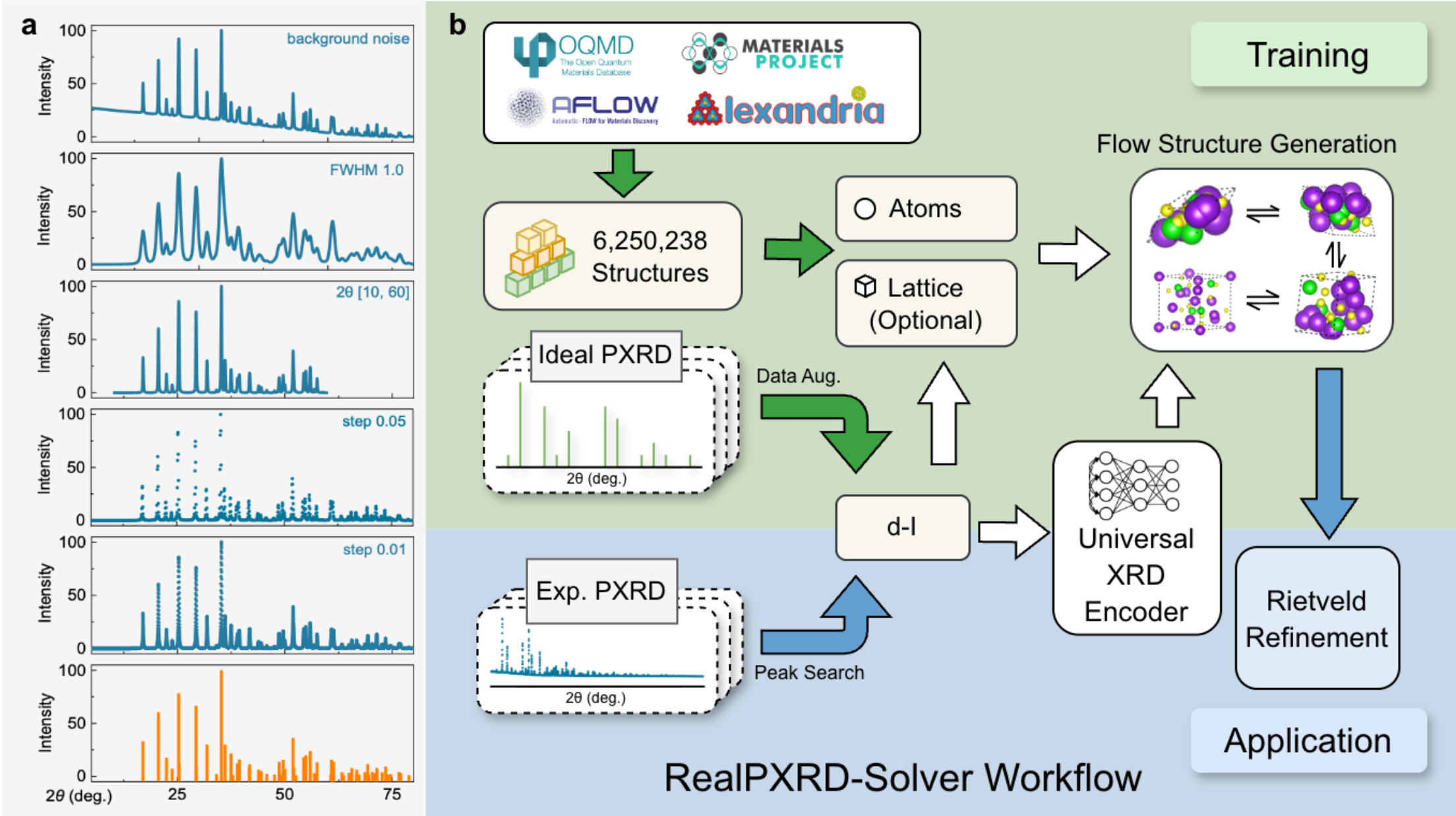

**Fig. 1. Design of RealPXRD-Solver for experimental PXRD data. (a)** Experimental variability in PXRD patterns of $LiFePO_4$. Patterns measured under different conditions (background noise level, peak width/FWHM, $2\theta$ range, and step size) nevertheless collapse to a consistent d–I (interplanar spacing–intensity) fingerprint, highlighting its invariance to measurement settings and sample quality. **(b)** Schematic of the full RealPXRD-Solver workflow. Simulated (training) and preprocessed experimental (inference) diffraction profiles are both converted to d–I lists and encoded by a universal XRD encoder into latent diffraction features. Conditioned on the chemical formula (and, when available, unit-cell parameters), a flow-based generative module proposes candidate crystal structures, which are subsequently ranked and refined by automated Rietveld analysis (e.g., GSAS-II).

Building on this invariant representation, RealPXRD-Solver couples the d–I input with a universal XRD encoder and a flow-based crystal structure generator (**Fig. 1b**). Both theoretical training patterns and preprocessed experimental patterns are transformed into d–I lists and passed through the same encoder, ensuring that the model learns a common latent description of diffraction features across simulated and real domains. The resulting latent diffraction features, together with the chemical formula, drive a conditional generative model that proposes candidate crystal structures; when unit-cell parameters are available from separate indexing or Le Bail analysis, they are incorporated as auxiliary conditioning information, reflecting standard crystallographic practice, while the model also retains a lattice-free ab initio setting for cases where indexing fails or the unit cell is unavailable. By aligning both the data representation and the generative workflow for simulated and experimental PXRD, this design directly addresses the simulation-to-reality gap highlighted in the Introduction and underpins the strong performance reported in subsequent sections.

### 2.2 RealPXRD-Solver achieves high accuracy on a large-scale theoretical benchmark

To assess the intrinsic capacity of RealPXRD-Solver independent of experimental imperfections, we first evaluated it on a large-scale benchmark of simulated PXRD patterns. In contrast to previous PXRD AI models such as Crystalyze[18] and PXRDnet[17], which are typically trained and tested on subsets of the Materials Project (for example, the MP-20 benchmark with approximately 45,000 structures and at most 20 atoms per cell)[25], RealPXRD-Solver is trained on a corpus of 6,250,238 unique crystal structures collected from multiple computational materials databases (**Fig. 2a-c** and **Supplementary Table 1**). This corpus spans all 7 crystal systems, 228 space groups, 89 chemical elements, and includes both simple and long-tail large-unit-cell structures, providing a much broader and more realistic prior over crystal chemistry than earlier benchmarks.

Consistent with the workflow-aligned design described above, we evaluate RealPXRD-Solver under two complementary settings on this theoretical benchmark. In the lattice-free setting, the model receives only the *d–I* fingerprint and chemical formula, corresponding to *ab initio* structure solution when unit-cell parameters are unavailable. In the lattice-conditioned setting, primitive-cell parameters from the reference structure are provided as auxiliary conditioning, reflecting scenarios in which reliable unit-cell information is obtained from indexing or Le Bail analysis. Importantly, reference CIFs are used exclusively for post hoc evaluation and are never provided to the model as inputs.

On a held-out theoretical test set of 10,000 structures drawn from this corpus (**Supplementary Fig. 2**), RealPXRD-Solver achieves high structure-recovery accuracy from their simulated PXRD patterns. In the lattice-free setting, the model attains match rates of 81.7% with a single generated candidate and 96.6% when considering 20 candidates per input. When primitive-cell parameters are additionally provided, performance further improves to 90.8% (Top-1) and 98.3% (Top-20), with an average structural root-mean-square error (RMSE) of approximately 0.02 across the benchmark (**Fig. 2d**). These results establish a strong upper bound on model performance under noise-free, label-consistent conditions, indicating that the remaining challenges in practical deployment arise primarily from experimental variability rather than limitations in generative capacity.

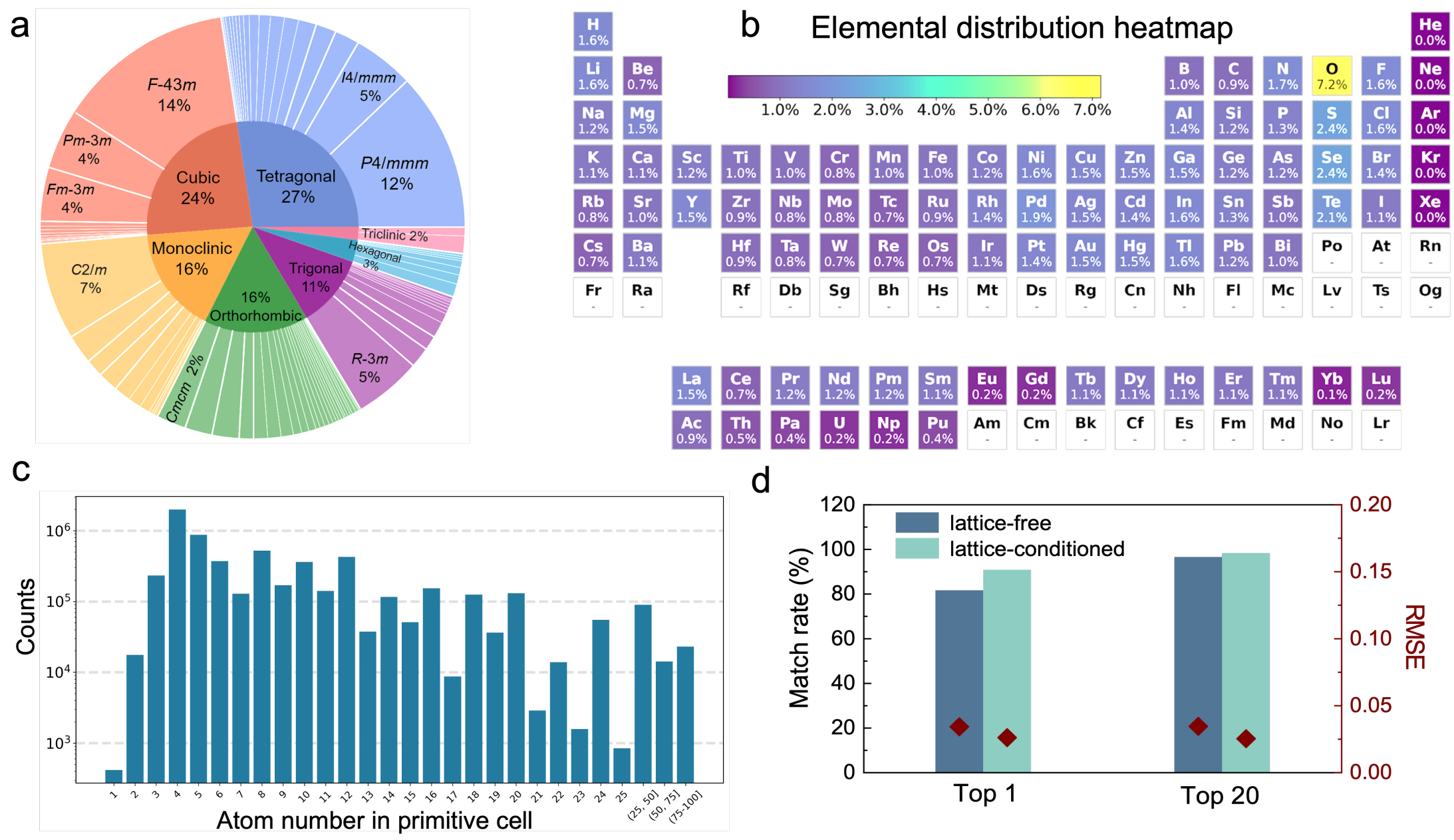

**Fig. 2. Dataset diversity and theoretical benchmark performance of RealPXRD-Solver. (a)** Distribution of crystal systems and space groups in the full theoretical corpus (6,250,238 unique entries). **(b)** Elemental occurrence heat map across the same corpus. **(c)** Distribution of primitive-cell atom counts, highlighting the long-tail extension beyond 25 atoms. **(d)** Cumulative Top-k match rates (Top-1 and Top-20) and mean structural RMSE (root-mean-square error) on the 10,000-structure test set, with and without lattice parameters conditioning.

### 2.3 Robustness against diverse experimental perturbations

While the large-scale theoretical benchmark establishes the intrinsic capacity of RealPXRD-Solver under idealized conditions, its practical value ultimately depends on robustness to the non-ideal features that pervade real PXRD measurements. In most real-world PXRD databases, including the PDF and COD, unit-cell parameters are routinely reported even when atomic coordinates are unavailable; accordingly, the following experimental case studies are evaluated in the lattice-conditioned setting unless otherwise stated. To probe robustness under such realistic conditions, we examined a set of challenging experimental patterns drawn from the Powder Diffraction File (PDF)[26] and COD[27], deliberately selecting cases with strong background contributions, pronounced preferred orientation, or impurity phases (**Fig. 3**).

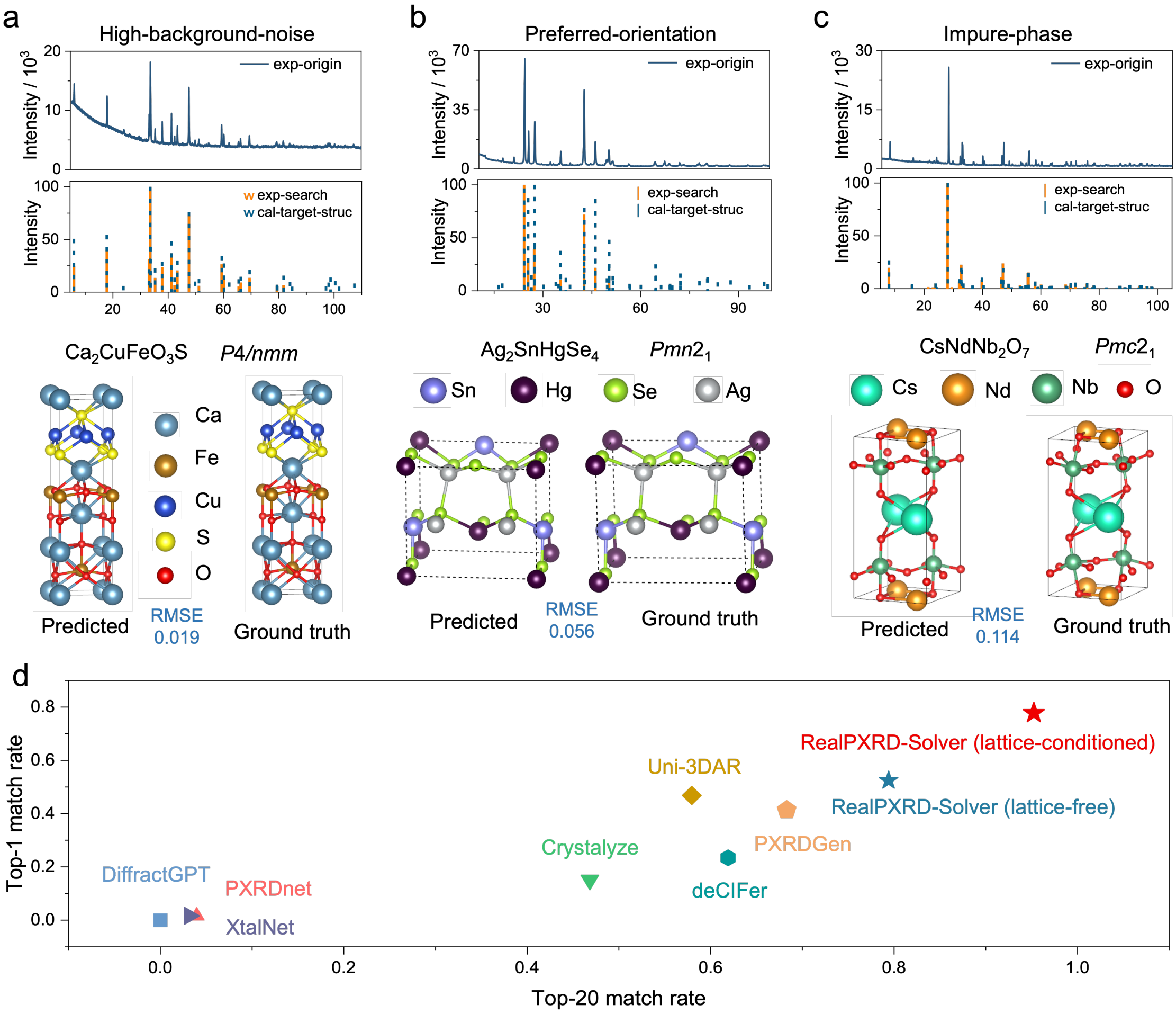

**Fig. 3. Robustness to experimental perturbations.** Multi-panel examples from the PDF database, showing input patterns (top), fitted vs. observed profiles (middle), and generated/refined structures (bottom). **(a)** High-background noise: $Ca_2CuFeO_3S$ (RMSE = 0.019). **(b)** Preferred orientation: $Ag_2SnHgSe_4$ (RMSE = 0.056). **(c)** Impurity phases: $CsNdNb_2O_7$ (RMSE = 0.114). All structures are displayed in primitive cells, with atoms color-coded by element. **(d)** Cross-model comparison on the CNRS subset of the opXRD experimental PXRD database. Scatter plot of Top-1 versus Top-20 structure match rates on 126 CNRS patterns for RealPXRD-Solver and previously reported generative PXRD-solving models.

High background noise, for example from air scattering or sample fluorescence, can obscure low-angle peaks and distort intensity ratios. For $Ca_2CuFeO_3S$ (PDF 00-062-0722, **Fig. 3a**), a laboratory pattern with substantial background, RealPXRD-Solver nonetheless recovered the correct tetragonal phase (space group *P*4/*nmm*) with a final structural root-mean-square error (RMSE) of 0.019 after Top-20 solution. This illustrates that the *d*–*I* encoding, combined with the learned diffraction prior, can effectively ignore broad background variations and focus on the underlying Bragg features.

Preferred orientation leads to large, systematic distortions in relative peak intensities and is a notorious failure mode for traditional structure-solution pipelines. In the case of $Ag_2SnHgSe_4$ (PDF 00-055-0291, **Fig. 3b**), where intensity deviations exceed 40% relative to the ideal pattern, RealPXRD-Solver still identified the correct orthorhombic structure ($Pmn2_1$), achieving an RMSE of 0.056. This example highlights that the model can tolerate severe intensity anisotropy as long as peak positions and overall motif remain recognizable.

Impurity and multiphase samples introduce additional peaks that are not explained by the target phase and can confound both indexing and refinement. For $CsNdNb_2O_7$ (PDF 00-055-0783, **Fig. 3c** and **Supplementary Fig. 3**), editorial comments indicate the presence of significant unindexed reflections, consistent with coexisting impurity phases. Despite these spurious peaks, RealPXRD-Solver, conditioned on the correct chemical formula and *d–I* fingerprint, isolated the primary phase and generated a structure in space group $Pmc2_1$ with an RMSE of 0.114.

Taken together, these case studies demonstrate that RealPXRD-Solver is able to reconstruct correct crystal structures under a wide range of experimental perturbations, including high background, strong preferred orientation, and impurity peaks. The model behaves as a robust, data-driven prior over physically plausible structures that, when coupled with automated refinement, can disentangle complex experimental signatures that are difficult to handle with purely physics-based methods alone.

### 2.4 Performance on quantitative experimental benchmarks

Beyond individual case studies, we quantitatively evaluated RealPXRD-Solver on two curated experimental benchmarks spanning diverse data-collection conditions and sample types: the CNRS experimental PXRD dataset from the opXRD database[23] and the RRUFF mineral PXRD database[24]. After applying preprocessing (see Methods, Data collection and preprocessing), the final evaluation sets contained 126 CNRS entries and 269 RRUFF entries, each paired with a well-refined reference CIF structure. In addition to these full evaluation sets, we also constructed structurally non-overlapping subsets comprising 23 CNRS entries and 14 RRUFF entries that have no symmetry-equivalent counterparts in the theoretical training corpus under the structure-matching criteria described in the Methods section, specifically the Evaluation metrics subsection. Unless otherwise specified, all experimental benchmark results reported below correspond to the lattice-conditioned setting, in which unit-cell parameters provided by the experimental metadata are supplied as auxiliary conditioning. This choice reflects standard experimental practice and enables a fair assessment of structure-solution performance given the information typically available to crystallographers. For completeness, we additionally report lattice-free results as a strict ab initio baseline for scenarios in which indexing fails or unit-cell parameters are unavailable.

On the CNRS dataset, which includes complex inorganic and mixed-metal oxides collected under well-controlled conditions, the model reached a Top-1 match rate of 77.9% and a Top-20 match rate of 91.9%, with a mean structural RMSE of 0.083 for the Top-1 candidate. On the RRUFF benchmark, which features mineralogical samples with fluorescence, background contamination, and variable texture, RealPXRD-Solver achieved a Top-1 match rate of 78.8% and a Top-20 match rate of 92.9%, with a mean structural RMSE of 0.102 for the Top-1 candidate. These results are summarized in **Table 1**, together with the performance on the structurally non-overlapping CNRS and RRUFF subsets.

**Table 1. Performance of RealPXRD-Solver on curated experimental PXRD benchmarks.** Results are reported for the full datasets (assessing simulation-to-reality robustness) and for structurally non-overlapping subsets (assessing generalization to structurally distinct entries), defined as experimental entries with no symmetry-equivalent counterpart in the theoretical training corpus under the structure-matching criteria (the Evaluation metrics subsection of Methods).

| Dataset | Number of entries | Top-1 match rate (%) | Top-20 match rate (%) | Mean structural RMSE (Top-1) |
| --- | --- | --- | --- | --- |
| CNRS (full set) | 126 | 77.9 | 91.9 | 0.083 |
| CNRS (non-overlapping subset) | 23 | 47.8 | 87.0 | 0.159 |
| RRUFF (full set) | 269 | 78.8 | 92.9 | 0.102 |
| RRUFF (non-overlapping subset) | 14 | 57.1 | 85.7 | 0.229 |

On the structurally non-overlapping subsets, which explicitly exclude any experimental entry with a symmetry-equivalent counterpart in the theoretical training corpus, RealPXRD-Solver is evaluated under the same lattice-conditioned setting as the full benchmark, but without access to any closely related structural exemplars during training. This regime therefore probes the model's ability to generalize to genuinely unseen crystal structure types, rather than to interpolate among memorized prototypes. While the Top-1 match rates decrease as expected given the absence of closely related structural precedents in the training corpus (47.8% for CNRS and 57.1% for RRUFF), the Top-20 match rates remain robustly high (87.0% and 85.7%, respectively). This indicates that even for crystallographic phases absent from the training distribution, the model reliably places the correct structure within a narrow candidate list. The disparity between Top-1 and Top-20 performance in this regime suggests that while the model successfully identifies the correct structural motif, the subsequent automated Rietveld refinement step, as integrated in our pipeline, is essential to unambiguously discriminate the optimal solution from structurally similar candidates.

Notably, the Top-20 accuracies of 91.9% on CNRS and 92.9% on RRUFF on the full evaluation sets approach the 98.3% Top-20 rate observed on the large-scale theoretical benchmark, indicating only modest degradation when transitioning from ideal simulated patterns to real experimental data. Even on the structurally non-overlapping subsets, Top-20 accuracies remain above 85%, suggesting that the main difficulties in experimental PXRD structure solution arise from a relatively small subset of particularly pathological patterns, rather than a systematic breakdown of the learned prior.

To place these numbers in the context of previously reported PXRD-solving approaches, **Fig. 3d** summarizes the Top-1 and Top-20 match rates on the CNRS benchmark for RealPXRD-Solver and a set of recent generative baselines, including PXRDnet[17], Crystalyze[18], PXRDGen[21], XtalNet[20], DiffractGPT[19], deCIFer[28], and Uni-3DAR[29]. Each point in the scatter plot corresponds to one model, evaluated on the same 126-entry CNRS subset and allowed to generate up to 20 candidates per pattern under its recommended inference setting (see the Methods section, specifically the Baseline models and CNRS benchmark protocol subsection). Under the lattice-conditioned setting, RealPXRD-Solver attains a Top-1 match rate of 77.9% and a Top-20 match rate of 91.9% on this benchmark, whereas previously reported generative models cluster at substantially lower accuracies in both coordinates (**Supplementary Table 2**). Importantly, RealPXRD-Solver also remains strong in the more challenging *ab initio* (lattice-free) regime, which removes unit-cell information and is therefore more directly comparable to many prior pipelines, achieving 52.4% Top-1 and 79.4% Top-20 on the same CNRS subset. In the Top-1 versus Top-20 plane, methods trained primarily on simulated PXRD patterns without explicit experimental augmentations tend to populate the lower-left region, reflecting a marked drop in accuracy when transferred to real experimental data, while RealPXRD-Solver lies near the upper-right corner, combining high single-candidate accuracy with strong Top-20 recall.

To further examine robustness to peak-picking non-idealities and experimental artifacts, we performed controlled perturbations on the CNRS benchmark peak lists, including weak-peak dropping (10-40%), near-peak merging ($\epsilon \leq 0.20^{\circ}$), peak-position jitter ($\pm 0.20^{\circ}$), and spurious peak insertion (1-5 peaks). As summarized in **Supplementary Table 3**, RealPXRD-Solver remains highly robust under the lattice-conditioned setting, with Top-1 match rates between 73.0% and 84.1% and Top-20 recall staying within 89.7-96.0%. Notably, this resilience extends to the more demanding *ab initio* (lattice-free) regime, where the model must simultaneously infer the unit cell and atomic positions. In this setting, Top-1 accuracies are maintained between 46.0% and 57.1%, while the Top-20 rates across most perturbation scenarios remain on par with the baseline performance (79.4%), even reaching 86.5% under spurious peak insertion. The stability of these metrics—particularly the model's ability to tolerate peak-position jitter and filter out false detections without prior lattice information—demonstrates that the learned crystallographic priors are sufficiently expressive to reliably recover structures even from significantly degraded experimental data. A full breakdown by perturbation type and setting is provided in **Supplementary Table 3**.

Taken together with the numerical comparison in **Supplementary Table 2**, these observations suggest that large, compositionally diverse training corpora and physically informed diffraction augmentations are key for robust generalization to experimental PXRD data. **Supplementary Table 4** further provides a side-by-side qualitative comparison with traditional diffraction-based structure-determination workflows (SXRD, conventional PXRD, and PXRD assisted by RealPXRD-Solver), illustrating the potential reductions in expert intervention and analysis time while retaining the ability to determine previously unknown structures from powder samples. Establishing a unified community benchmark with standardized evaluation protocols across models and datasets would be a valuable direction for future work.

### 2.5 Overcoming inherent limitations of powder X-ray diffraction

Beyond common experimental artifacts, many of the most difficult PXRD problems stem from intrinsic limitations of X-ray scattering itself. These include distinguishing neighboring elements with similar scattering factors, locating light atoms such as hydrogen, and solving structures with large primitive cells where peak overlap and parameter degeneracy become severe. We used representative cases from the PDF and CNRS databases to probe how far RealPXRD-Solver can push beyond these traditional limits (**Fig. 4**). Given that these database entries typically provide experimental unit-cell parameters derived from prior indexing, we utilized the model in its lattice-conditioned mode. This allows the structural prior to focus specifically on the spatial arrangement of atoms within a fixed chemical and geometric framework, mirroring the standard workflow for resolving “coordinate-less” entries.

A fundamental challenge in PXRD is the difficulty in precisely locating certain atoms within a crystal structure. This includes neighboring elements on the periodic table (e.g., Co, $Z$=27 vs. Mn, $Z$=25), which produce minimal alterations in diffraction patterns, and light elements like hydrogen, which contribute negligibly to X-ray scattering. By leveraging knowledge of stable elemental configurations learned from extensive datasets, RealPXRD-Solver can overcome these challenges. For $Ca_3CoMnO_6$ (PDF 04-006-8682, rhombohedral $R$-3$c$, **Fig. 4a**), the model correctly assigned the Co/Mn sites from a lab-collected pattern, yielding a structure that aligns with neutron-validated coordinates. The final result achieved an RMSE of 0.025. In the case of $MnPO_4 \cdot H_2O$ (PDF 00-051-1548, monoclinic $C2/c$, **Fig. 4b**), a compound whose structure was previously determined using synchrotron data, RealPXRD-Solver successfully generated a similar structure with an RMSE of 0.100, accurately accounting for the presence of hydrogen.

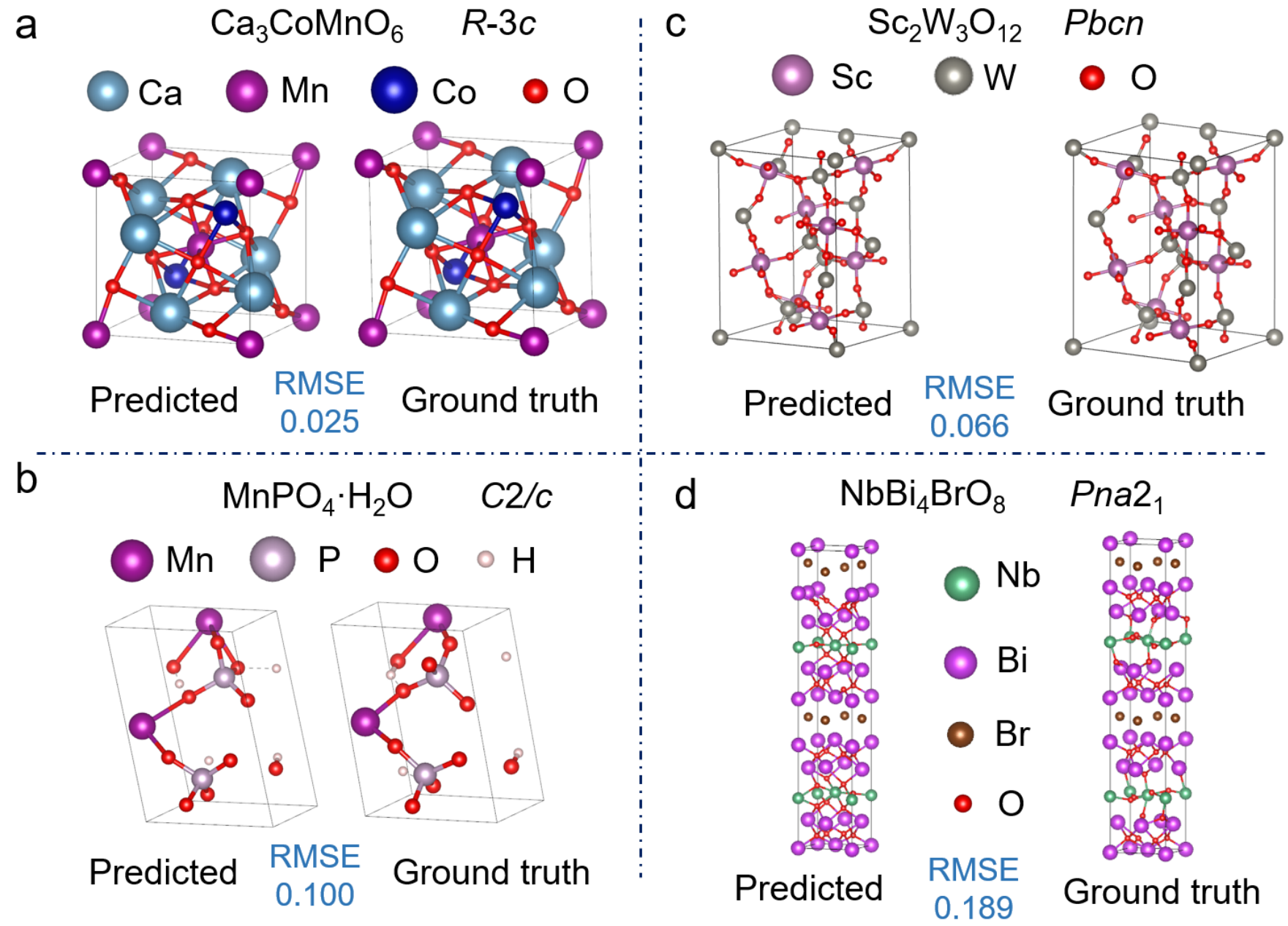

**Fig. 4. Overcoming challenging scenarios in PXRD structure determination.** Examples from the PDF and CNRS datasets, comparing predicted and observed primitive structures. **(a)** Distinguishing neighboring elements: $Ca_3CoMnO_6$, correct Co/Mn assignment (RMSE = 0.025). **(b)** Locating light atoms: $MnPO_4{\cdot}H_2O$ (RMSE = 0.100). **(c, d)** Solving large unit cells: **(c)** $Sc_2W_3O_{12}$ (68 atoms, RMSE = 0.066). **(d)** $NbBi_4BrO_8$ (56 atoms, RMSE = 0.189).

Another significant challenge involves solving structures with a high number of atoms in the primitive cell. The performance of the base RealPXRD-Solver model can be limited in this area due to the long-tail distribution of its training data, in which the vast majority of crystal structures contain fewer than 25 atoms. To address this, we implemented a strategy of retraining the model on a specialized dataset. We curated a long-tail dataset containing only structures with more than 25 atoms and used it to fine-tune the base model. This newly weighted model showed substantially improved performance on complex structures. For example, when applied to $Sc_2W_3O_{12}$ (68 atoms) and $NbBi_4BrO_8$ (56 atoms), the base model failed to generate structures matching the ground truth. In contrast, the fine-tuned long-tail model successfully generated accurate structures for both, as illustrated in **Fig. 4c and d**.

### 2.6 Automated solution of unreported structures from the PDF database

A hallmark of RealPXRD-Solver's utility is its ability to solve crystal structures that have historically resisted solution. Diffraction databases like the Powder Diffraction File (PDF) contain hundreds of thousands of such "coordinate-less" entries. These entries typically provide well-defined unit-cell parameters derived from indexing but lack resolved atomic coordinates, making them directly amenable to our model's lattice-conditioned mode. Automatically solving and refining these unknown structures is critical for discovering novel materials in future AI-driven robotic laboratories[22], yet it has traditionally required weeks of expert-led, trial-and-error analysis.

To address this challenge efficiently, we coupled RealPXRD-Solver to an automated, end-to-end structure-solution pipeline (see the Methods section, specifically the automated pipeline for solving coordinate-less PDF entries subsection, and **Supplementary Fig. 4**). Starting from reported unit cells and experimental PXRD patterns, the pipeline generates structural candidates, refines them under symmetry and chemical constraints, and evaluates their agreement with the data based on standard Rietveld reliability factors.

Considering that the PDF license does not permit large-scale benchmarking over its entire database, we validate our pipeline by applying it to a selection of unsolved entries. Notably, this approach successfully determined the crystal structures of 39 materials whose crystal structures had remained unknown since their initial reporting. Representative examples are shown in **Fig. 5** and **Supplementary Fig. 5**, and all solved structures are provided in the Extended Data.

## 3. Conclusions

RealPXRD-Solver provides a robust and scalable framework for determining crystal structures directly from experimental PXRD data. By combining an invariant *d–I* (interplanar spacing-intensity) fingerprint, a universal XRD encoder shared between simulated and experimental domains, and a flow-based generative backbone, the method effectively bridges the long-standing simulation-to-reality gap. Unlike previous AI solvers that conflate indexing and structure solution, our workflow-aligned design supports both lattice-conditioned and lattice-free (*ab initio*) modes. This flexibility allows the model to leverage reliable unit-cell metadata when available, mirroring standard crystallographic practice, while maintaining high performance in scenarios where indexing is unavailable.

Our results demonstrate that large-scale training on over 6.2 million theoretical structures, augmented by physics-informed experimental perturbations and targeted long-tail fine-tuning for complex cells, enables genuine generalization. This is evidenced not only by the near-perfect recovery (98.3% Top-20) on theoretical benchmarks but also by Top-20 accuracies exceeding 90% on two curated experimental datasets (CNRS and RRUFF). Notably, the model's performance remains robustly above 85% even on structurally non-overlapping subsets, indicating that it has learned deep crystallographic priors rather than merely interpolating between memorized prototypes. Furthermore, the model's ability to tolerate peak-position jitter, noise, and impurity phases—and its capacity to resolve neighboring elements and light atoms—establishes it as a reliable tool for practical structure determination.

The real-world utility of RealPXRD-Solver is most clearly illustrated by its integration into an automated pipeline that successfully solved 39 previously unreported structures from the PDF database. This demonstration marks a transition from proof-of-concept AI modeling to a functional engine for high-throughput phase analysis. As summarized in **Supplementary Table 4**, our framework significantly reduces the need for expert intervention and specialized single-crystal or synchrotron measurements, offering a rapid, automated alternative for routine laboratory PXRD data.

Despite these advances, limitations remain regarding the representation of organic-hybrid systems, highly disordered materials, and complex multiphase mixtures. Future work will focus on expanding the training distribution to include these regimes and integrating RealPXRD-Solver into autonomous laboratory platforms to enable closed-loop materials discovery. We anticipate that the principles of invariant representation and workflow-aligned generative modeling demonstrated here will provide a blueprint for a wide range of diffraction-based characterization tasks in the era of AI-driven science.

# 4. Methods

### 4.1 Data collection and preprocessing

The RealPXRD-Solver workflow integrates two parallel data pipelines, one for theoretical diffraction patterns and another for experimental PXRD data, to ensure consistent representation between simulated and real-world inputs.

***Theoretical data.***

Theoretical crystal structures were aggregated from several large-scale computational materials databases, including AFLOW[30], Alexandria[31, 32], the Materials Project[25], and OQMD[33, 34] (see Supplementary Table 1 for a detailed breakdown). Starting from these raw collections, we first enforced basic thermodynamic stability criteria (energy above the convex hull < 0.1 eV/atom where available) and then removed duplicate or symmetry-equivalent entries using Pymatgen's structure matcher[35] with fixed tolerances (stol = 0.5, angle_tol = 10°, ltol = 0.3). After filtering and deduplication, the final theoretical corpus contained 6,250,238 unique structures spanning all 7 crystal systems, 228 space groups, and 89 chemical elements, with primitive-cell atom counts ranging up to 444 (Fig. 2a-c).

For each entry in this corpus, the powder X-ray diffraction pattern was simulated using the Pymatgen XRD module[35] under Cu~Kα radiation ($\lambda$ = 1.5406 Å) over a 2θ range of 5°-80°. The resulting discrete *d-I* (interplanar spacing-intensity) lists are normalized by their maximum intensity to match the input format used for experimental data. These simulated d-I lists constitute the theoretical branch of the training and evaluation pipelines. For the simulated-data evaluation, we constructed a held-out test set consisting of 10,000 crystal structures randomly sampled from the 6,250,238-entry theoretical corpus (see the Evaluation metrics subsection of Methods for details).

***Experimental PXRD patterns.***

Experimental PXRD patterns were collected from three primary sources: the CNRS experimental PXRD dataset from the opXRD database[23], the RRUFF mineral diffraction database[24], and the Powder Diffraction File (PDF)[26]. Among these, CNRS and RRUFF were used for quantitative benchmarking, whereas PDF entries were primarily employed as case studies for automated solution of previously unreported structures.

For the CNRS subset, we similarly retained only inorganic crystalline phases with refined CIFs and primitive-cell atom counts not exceeding 20, explicitly excluding purely organic and hybrid materials to avoid compositional regimes underrepresented in the training corpus and to remain within the input regime handled by most baseline methods. For RRUFF, we selected mineral entries that (i) have an associated CIF structure deemed reliable by the database curators, and (ii) contain no more than 20 atoms in the primitive cell. This cutoff follows the widely adopted MP-20 setting and ensures a fair comparison with existing PXRD-solving models, many of which are designed and reported only for structures with up to 20 atoms per primitive cell.

To standardize the scattering geometry, all experimental patterns were converted to an equivalent Cu~Kα wavelength: reported diffraction profiles were mapped to d-spacing and then re-expressed on a Cu~Kα ($2\theta$) scale.

Finally, 126 CNRS patterns and 269 RRUFF patterns remained and were used for quantitative evaluation.

To further assess generalization beyond any potential structural overlap, we also constructed structurally non-overlapping subsets of the CNRS and RRUFF benchmarks. For each experimental entry, the reference CIF structure was compared against all theoretical training structures using the same Pymatgen structure matcher and fixed tolerances (stol = 0.5, angle_tol = 10°, ltol = 0.3) described below. Entries that did not admit any symmetry-equivalent counterpart in the training corpus under these criteria were retained in the non-overlapping subsets, resulting in 23 CNRS and 14 RRUFF patterns.

All retained experimental patterns (CNRS, RRUFF, and PDF) were subjected to a uniform preprocessing protocol. Smoothing and baseline correction were performed using polynomial background subtraction (order 4–6), followed by local-maxima detection to identify Bragg peaks. The resulting discrete d–I lists were normalized by maximum intensity and truncated or padded to a fixed sequence length for batching.

This d–I representation serves as a stable structural fingerprint, inherently more invariant than raw intensity-$2\theta$ profiles to experimental variations such as background level, counting noise, peak broadening, $2\theta$ range, and step size. The consistency between experimental and simulated d–I distributions is shown in Supplementary Fig. 1.

### 4.2 Data augmentation for experimental variability

To emulate the variability of real PXRD measurements, theoretical patterns were simulated using the Pymatgen diffraction module (Cu Kα, $2\theta$ range: 5°–80°) and subsequently augmented with synthetic perturbations. The augmentation pipeline included: (i) Random peak shifts of ±0.1° in $2\theta$ to mimic instrumental offsets; (ii) Gaussian noise addition $\sigma = 0.05$ to reproduce counting statistics and detector noise; (iii) Intensity scaling factors (0.8–1.2) to simulate preferred orientation or partial texture effects; (iv) Minor peak removal or broadening to emulate incomplete crystallinity or strain. These procedures ensure that the model encounters diffraction data representative of diverse experimental conditions, bridging the simulation-to-reality distribution gap during training.

### 4.3 Model architecture: Universal XRD encoder

The *d–I* list serves as the fundamental feature set for a PXRD pattern. Our encoder operates directly on this list by treating it as a sequence of $n$ peaks, characterized by their continuous d-spacing values $\{d_1, ..., d_n\}$ and corresponding intensity values $\{I_1, ..., I_n\}$. These two scalar inputs are first projected into high-dimensional feature vectors, denoted as $f_{d,i}$ and $f_{I,i}$, using two independent multilayer perceptrons (MLPs):

$$f_{d,i}=MLP_d(d_i), \quad f_{I,i}=MLP_I(I_i)$$

To obtain a single representation for each peak, we sum these two vectors element-wise:

$$f_{peak,i}=f_{d,i}+f_{I,i}, \quad i=1,...,n,$$

resulting in a unified peak feature sequence $F_{peaks}=\left(f_{peak,1},...,f_{peak,n}\right)$.

Following standard practice for sequence modeling, a learnable global [*CLS*] token (denoted as $f_{cls}$) is prepended to the sequence to serve as an aggregator for the global pattern representation:

$$F_{in}=Concat\left(\left[f_{cls}, F_{peaks}\right]\right)$$

The input sequence $F_{in}$ is then passed through a multi-layer Transformer encoder, adapted from the XtalNet architecture[20], which models contextual dependencies among all peaks. We use the final hidden state corresponding to the [*CLS*] token as the holistic latent representation $F_{XRD}$ of the entire PXRD pattern.

In practice, $F_{XRD}$ is projected to a 512-dimensional latent vector, which is concatenated with auxiliary conditioning information, including an embedding of the chemical formula and, when available, a compact encoding of the unit-cell parameters. This fused conditioning vector is then fed into the downstream flow-based crystal structure generator to guide the joint prediction of lattice parameters and atomic coordinates in the primitive cell.

### 4.4 Model architecture: Flow-based structure generator

The generative backbone follows a flow-matching architecture inspired by DiffCSP[36] and FlowMM[37].

The model jointly predicts the lattice matrix ($L$) and fractional atomic coordinates ($\{F_j\}$) in the primitive cell by learning their deterministic flow fields.

$$L_t=(1-t)L_0+tL_r$$

$$F_t=F_0+tF_r$$

And the generative process follows,

$$L_{t-\Delta t}=L_t-\frac{L_r-L_t}{1-t}\Delta t$$

$$F_{t-\Delta t}=F_t-(1+5t)F_r\Delta t$$

To predict the noise of the lattice ($L_r$) and atomic coordinates ($F_r$), message passing mechanism is modeled as,

$$\hat{\epsilon}_L,\hat{\epsilon}_F=GNN\left(L,\{F_j\},\{h_j\}\right)$$

where the GNN represents a graph neural network to embed the structures, and node features $h_j$ represent latent information of atom types, time and XRD features ($f_{XRD}$).

And the training objective is defined as,

$$L_{total}=\lambda_1|\hat{\epsilon}_L-L_r|^2+\lambda_2|\hat{\epsilon}_F-F_r|^2$$

where $\lambda_1$ and $\lambda_2$ are tunable weights.

### 4.5 Training strategy and hyperparameters

The AI model is implemented in PyTorch and PyTorch Lightning framework, and trained on four NVIDIA A100-PCIE-80GB GPUs for about one week due to the large dataset. The model was trained for 500 epochs, with a batch size of 480 for the full dataset and 64 for the long-tail dataset. We used the Adam optimizer with a learning rate of $1e^{-3}$, together with a ReduceLROnPlateau scheduler. The dataset was randomly split into 99% for training and 1% for validation. To improve performance on complex structures with > 25 atoms per primitive cell, the base model was fine-tuned on a curated long-tail dataset enriched in large-unit-cell compounds.

### 4.6 Inference and automatic structure refinement

At inference, each PXRD input yields 20 structural candidates. The generated crystal structures are automatically refined using GSAS-II[38] through a scripted Rietveld procedure that adjusts atomic positions and lattice parameters under chemical and symmetry constraints. A structure is considered successfully solved if it achieves an $R_{wp}$ value below 15% and displays geometric consistency with known coordination chemistry. The average inference time per sample is less than one minute on a single GPU.

### 4.7 Evaluation metrics

We report Top-*k* match rates based on structural similarity between predicted and reference CIF structures, using Pymatgen's structure matcher[35] with fixed tolerance (stol=0.5, angle_tol=10°, ltol=0.3) across all experiments. The structural RMSE, combining deviations in lattice parameters and fractional atomic coordinates, was used as a continuous accuracy metric. When multiple candidate structures were generated, the best-matching candidate for each pattern was used for RMSE calculation and for Top-*k* statistics. The same structure-matching configuration was used both for deduplicating the theoretical corpus and for defining the structurally non-overlapping subsets of the CNRS and RRUFF experimental datasets.

### 4.8 Peak-list perturbation robustness evaluation

To quantify robustness to errors introduced by peak picking and experimental imperfections, we performed controlled perturbations directly on the extracted discrete peak list while keeping the same CNRS evaluation set, the same inference hyperparameters, and the same structure-matching criteria.

In this evaluation, each experimental diffractogram is represented in the Cu Kα-equivalent ($2\theta$, $I$) peak-list form, where $2\theta$ is in degrees and intensities are normalized. Given the baseline peak list $\{(\theta_i, I_i)\}_{i=1}^{n}$, we applied one perturbation at a time:

(i) *Weak-peak dropping:* a fraction $\rho \in \{0.10, 0.20, 0.30, 0.40\}$ of the weakest peaks (lowest $I_i$) was removed, keeping the remaining top-$(1\text{-}\rho)$ peaks (at least one peak retained).

(ii) *Near-peak merging:* peaks were sorted by $\theta$, and consecutive peaks satisfying $\theta_{i+1}\text{-}\theta_i < \epsilon$ were merged using $\epsilon \in \{0.05^\circ, 0.10^\circ, 0.20^\circ\}$. Within each merged cluster $C$, the new intensity was set to $I^* = \sum_{j \in C} I_j$, and the new position was set to the intensity-weighted average $\theta^* = \left(\sum_{j \in C} \theta_j I_j\right) / \left(\sum_{j \in C} I_j\right)$.

(iii) *Peak-position jitter:* each peak position was perturbed as $\theta_i^{'}=\theta_i+u_i$ with $u_i \sim$ Uniform($-\delta, \delta$) and $\delta \in \{0.05°,0.10°,0.20°\}$.

(iv) *Fake-peak insertion:* $m\in\{1,3,5\}$ additional peaks were inserted with $\theta_{fake} \sim Uniform(min_i\theta_i, max_i\theta_i)$ and intensities sampled as $I_{fake}=r \cdot I_{max}$ *with* $r \sim Uniform(0.02,0.08)$.

After perturbation, peaks were sorted by $\theta$ and intensities were re-normalized to keep $max_i I_i =100$.

**4.9 Baseline models and CNRS benchmark protocol**

For the cross-model benchmark on experimental data shown in Fig. 3d and Supplementary Table 2, we compared RealPXRD-Solver against seven recently proposed generative approaches for PXRD-based crystal structure determination: PXRDnet[17], Crystalyze[18], PXRDGen[21], XtalNet[20], DiffractGPT[19], deCIFer[28] and Uni-3DAR[29]. All models were evaluated on the same CNRS subset of the opXRD database[23] described in the Methods section, specifically the Data collection and preprocessing subsection. In brief, this subset consists of 126 single-phase inorganic entries for which reliable CIF structures, chemical compositions, and unit-cell parameters are available, and whose primitive cells contain no more than 20 atoms. The 20-atom cutoff mirrors the MP-20 regime and is chosen to ensure that all baseline models can be applied under their originally reported settings, since several previously published PXRD-solving methods do not support, or have not been benchmarked on, structures with larger primitive cells.

To ensure that all methods operated on comparable inputs, we adopted a two-stage preprocessing strategy. First, all CNRS diffractograms were brought to a common scattering geometry: the raw intensity–$2\theta$ profiles were converted to d-spacing, mapped to an equivalent Cu~Kα wavelength, and, where necessary, re-expressed on a Cu~Kα $2\theta$ scale. The intensities were then normalized (by the maximum intensity) to remove arbitrary scale factors. Second, model-specific input formatting was applied following the original implementations. For RealPXRD-Solver, we extracted peak positions and intensities from the normalized patterns and constructed discrete d-I lists as described in the Methods section, specifically the Data collection and preprocessing subsection. For baseline models that require fixed $2\theta$ ranges and step sizes, the experimental patterns were interpolated onto a uniform $2\theta$ grid matching the model's prescribed angular range and step size, and then padded or truncated to the required input length. For methods with different input conventions, we followed the preprocessing pipelines recommended by the respective authors. Whenever a method supports conditioning on composition or unit-cell parameters, we provided the same chemical formula and primitive-cell information used by RealPXRD-Solver.

For each model and each CNRS pattern, we allowed the method to propose up to 20 candidate crystal structures under its recommended inference settings (number of samples, sampling schedule, and internal ranking strategy). Structural similarity to the CNRS reference CIFs was assessed with the same Pymatgen structure matcher and tolerance settings as described in the Methods section, specifically the Evaluation metrics subsection, ensuring that all methods were judged by a common structural criterion.

**4.10 Automated pipeline for solving coordinate-less PDF entries**

For coordinate-less entries in the PDF database, we implemented a fully scripted structure-solution pipeline built on top of RealPXRD-Solver and GSAS-II. The workflow proceeds as follows.

First, for each selected PDF entry, the reported unit cell is converted to its primitive form using the *NIST*LATTICE* software package[39]. The corresponding experimental PXRD pattern is preprocessed into a discrete d–I list using the same baseline-correction, peak-picking, and normalization procedures described in the Methods section, specifically the Data collection and preprocessing subsection. This *d–I* fingerprint, together with the chemical formula, is then used as the conditioning input to RealPXRD-Solver.

Second, RealPXRD-Solver generates 20 candidate crystal structures per PDF entry, all expressed in the primitive cell. These candidates are ranked by comparing the agreement between their simulated PXRD patterns (computed with Pymatgen under Cu~Kα radiation) and the experimental d–I list.

Third, the top-ranked candidates are passed to a scripted GSAS-II[38] Rietveld refinement routine. In this step, lattice parameters and fractional atomic coordinates are refined under the reported space-group symmetry, with additional bond-length and bond-angle restraints imposed where applicable. Background, peak-profile, and scale parameters are also optimized to obtain a stable fit.

Finally, a structure is considered successfully solved if the refined model achieves a weighted-profile R factor $R_{wp}$ below 15% and yields a visually satisfactory fit to the experimental pattern, without unindexed major reflections.

## Supplementary Information

Supplementary Information is available from https://doi.org/x-x.

## Acknowledgments

The authors acknowledge financial support from National Natural Science Foundation of China (Grant No. 52272268, 21721002) and Beijing Municipal Science & Technology Commission, Administrative Commission of Zhongguancun Science Park (Grant No. Z251100007525007). We thank our colleagues at DP Technology and the participating institutions for their valuable discussions and technical support throughout this project.

## Author contributions

Q.L., Z.Y., G.K., W.E., Z.T., S.J. and L.Y. conceived the project and designed the RealPXRD-Solver framework. Q.L., L.Y., and R.J. developed the core generative model and the universal XRD encoder. Q.L., L.Y., M.G., J.G., and H.X. performed large-scale data collection, deduplication, and augmentation. Q.L. and F.X. carried out model training. Z.Y. and Q.L. conducted the benchmarks on theoretical and experimental datasets (CNRS and RRUFF). Q.L. and M.G. developed the automated Rietveld refinement pipeline. Q.L. and L.Y. provided technical support for the long-tail structure analysis. Z.Y., G.K., W.E., Z.T., S.J., L.Y., W.Z., W.H., J.Y., L.Z., and C.W. provided strategic scientific guidance, contributed to the theoretical interpretation of the results, and supervised the research. Q.L., S.J., L.Y., Z.Y., and J.G. wrote the manuscript with input from all authors. All authors discussed the results and commented on the final version of the paper.

## Competing interests

The authors declare no competing interests.

## Data availability

All data and materials used in the analysis are available to any researcher for purpose s of reproducing or extending the analysis. The large-scale dataset, comprising approxi mately 6,250,238 unique theoretical entries and long-tail entries with more than 25 ato ms, along with the trained model weights, are available at https://drive.google.com/driv e/folders/1WHXuSx9LD5NVFeSkeQs3_AZ8G_ZBm6MX. The code for the RealPXRD-Solver is available at https://github.com/liqi-529/RealPXRD-Solver.git.

## Code availability

The code for the RealPXRD-Solver is available at https://github.com/liqi-529/RealPXR D-Solver.git.